# A National Research Agenda for Intelligent Infrastructure

| Elizabeth Mynatt | Jennifer Clark | Greg Hager |
| Georgia Tech | Georgia Tech | Johns Hopkins University |
| Computing Community Consortium | | |

Dan Lopresti — Lehigh Univeristy

Greg Morrisett — Cornell University

Klara Nahrstedt — University of Illinois

George Pappas — University of Pennsylvania

Shwetak Patel — University of Washington

Jennifer Rexford — Princeton University

Helen Wright — Computing Community Consortium

Ben Zorn — Microsoft Research

Our infrastructure touches the day-to-day life of each of our fellow citizens, and its capabilities, integrity and sustainability are crucial to the overall competitiveness and prosperity of our country. Unfortunately, the current state of U.S. infrastructure is not good: the American Society of Civil Engineers' latest report on America's infrastructure ranked it at a D+ — in need of $3.9 trillion in new investments. This dire situation constrains the growth of our economy, threatens our quality of life, and puts our global leadership at risk. The ASCE report called out three actions that need to be taken to address our infrastructure problem: 1) investment and planning in the system; 2) bold leadership by elected officials at the local and federal state; and 3) planning sustainability and resiliency in our infrastructure[1].

While our immediate infrastructure needs are critical, it would be shortsighted to simply replicate more of what we have today. By doing so, we miss the opportunity to create *Intelligent Infrastructure* that will provide the foundation for increased safety and resilience, improved efficiencies and civic services, and broader economic opportunities and job growth. Indeed, our challenge is to proactively engage the declining, incumbent national infrastructure system and not merely repair it, but to enhance it; to create an internationally competitive cyber-physical system that provides an immediate opportunity for better services for citizens and that acts as a platform for a 21st century, high-tech economy and beyond.

Intelligent infrastructure is the deep embedding of sensing, computing, and communications capabilities into traditional urban and rural physical infrastructures such as roads, buildings, and bridges for the purpose of increasing efficiency, resiliency, and safety. For example, embedding controllers, intersection schedulers, and sensors along roads creates new capabilities to control traffic signals and optimize traffic flow. Developing intelligent power grids provides the means to provide new efficiencies in power generation and transmission, improved resilience to natural and human-originated disruptions, and streamlined integration of new energy sources such as wind. These are all examples of how intelligence infrastructure will change the way we interact with each other and how we plan and design our cities, towns, and communities. It is essential for us as a nation to invest in intelligent infrastructure and establish a coordinated research agenda to reap the benefits of this revolutionary change.

Across disciplines ranging from engineering to computer science to public policy, intelligent infrastructures are increasingly seen as solutions to the long-standing problems that face local governments attempting to respond to both long term and short term threats to resilience: 1) strained resources spread across ever growing urban populations, 2) aging infrastructures and public services systems, 3) competitiveness in the global economy, and 4) acute human and environmental stressors due to rapid growth and change in regional areas.

---

[1] http://www.infrastructurereportcard.org/

The economic opportunity presented by intelligent infrastructure is three-fold. First, the data produced by intelligent infrastructure promises to increase the reliability of local government services and performance of infrastructure systems. For example, deep analysis of traffic patterns in connection with more intelligent traffic lights can significantly reduce traffic congestion on roadways, and smart meters can record consumption of electric energy and send that information back to a utility for monitoring and billing. The data also paves the way for building interoperable and cross platform systems that enable cross-system coordination, and that will ultimately allow localities to provide higher quality, expanded services, from waste management to emergency response to public health, at a lower cost.

The second opportunity is that intelligent infrastructure investments can enhance and inform the strategic planning capacities of local communities—large and small—with real-world data on how infrastructure is used by citizens and businesses and how the infrastructure is performing. Local communities, businesses, and citizens will be able to see how their community is operating, and receive prescriptive recommendations and predictive insights to inform future investments, ranging from immediate actions to counter flooding risks, to proactively addressing increasing risks to public safety due to accidents, crime or natural disasters.

Third, the sharing of data amongst smart community partners and participants helps to build networks for sharing policy strategies and technical approaches. These strategic partnerships form the foundation of economic opportunities that flow from far-reaching investments in intelligent infrastructure: entrepreneurship and market leadership. The data generated by and for smart community systems (and the systems that produce that data) form the foundation of new enterprises and new products and services and, as a consequence, function as platforms for further economic development.

The Federal Government has an important role to play in shaping the scope and scale of intelligent infrastructure investments going forward. After World War II, as the nation's infrastructure exploded to accommodate returning veterans and their families, the production of the Interstate Highway System began. This national project to invest in a critical grid system to cover our country revolutionized our nation. It forever changed how we transport our goods and services and it has allowed and contributed to the continued growth and expansion of our cities and towns. In short, the Federal Government will decide the platform on which the national economy is built going forward and whether it meets 20th century standards or sets a new standard for the 21st century economy.

How to design and deploy intelligent infrastructure to efficiently and effectively support our communities is one of the central questions going forward for the country as a whole, and for local communities in specific. Federal action, like the Federal Aid Highway Act, is needed to start a national project of this scale. It is critical that we obtain this support now as we start thinking about the potential of intelligent infrastructure and all its possible impacts in our daily lives.

**1.0    The Potential of Intelligent Infrastructure**

We define "Intelligent Infrastructure" as "the integrated sensing and data analytics with municipal capabilities and services that enable evidence-based operations and decision making" which is:

- **Descriptive**: Provides an accurate and timely characterization of current state, e.g., water level in a storm drain or traffic congestion.
- **Prescriptive**: Recommends immediate and near-term actions, e.g., re-routing traffic or dispatching onsite service personnel.
- **Predictive**: Anticipate future challenges and opportunities, based on assessment of the current state, patterns of past activity and available resources and capabilities, e.g., street-level flooding by incorporating water sensors, weather patterns and runoff capabilities.
- **Proactive**: Guides complex decision making and scenario planning, incorporating economic data, to inform future investment.

Intelligent Infrastructure can **incrementally increase operational performance and the ability to guide improvements through decision support**. These "loops" or cycles of learning span automation and decision support to the eventual production of generalized knowledge. For example, advanced transportation systems could incrementally learn to manage different patterns of traffic, then provide decision support for managing special cases (e.g., disaster response), then support planning and prioritization for new road/control modifications, and finally document generalized knowledge that can be applied across different cities with varying transportation capabilities. Intelligent Infrastructure has the potential to transform daily life and civic services across many configurations of municipal systems and services (see Table 1).

**Table 1: The Potential of Intelligent Infrastructure Across Multiple Domains**

| | **Descriptive** | **Prescriptive** | **Predictive** | **Proactive** |
|---|---|---|---|---|
| **Intelligent Transportation** | Real time traffic congestion information | Reroute traffic; Adjust dynamic lane configuration (direction, HOV) | Anticipate rush hour / large event congestion; Anticipate weather related accidents | Suggest traffic patterns w/ intelligent stoplights; road diet plan |
| **Intelligent Energy Management** | Real time energy demand information | Improve asset utilization and management across transmission and distribution system | Anticipate demand response required to ensure grid reliability | Suggest new market approaches to integrate production and distribution capabilities |
| **Intelligent Public Safety and Security** | Real time crowd analysis | Threat detection; Dispatch public safety officers | Anticipate vulnerable settings and events | Suggest new communication and coordination response approaches |
| **Intelligent Disaster Response** | Real time water levels in flood prone areas | Timely levee management and evacuations as needed | Anticipate flood inundation with low-cost digital terrain maps | Inform National Flood Insurance Program; Inform vulnerable populations |
| **Intelligent City Systems** | Describe mobility patterns (pedestrian, cycling, automobile, trucking, electric and autonomous vehicles) | Adjust mobility management to improve safety; reduce energy usage | Anticipate changing needs for parking, charging stations, bike and ride share programs. | Inform future mobility capabilities to drive economic development and reduce barriers to employment |
| **Intelligent Agriculture** | Characterize spatial and temporal variability in soil, crop, and weather. | Advise based on environmental stressors and crop traits | Forecast crop yield; Anticipate seasonal water needs | Customize management practices and seed selection to local conditions |
| **Intelligent Health** | Block – level assessment of current allergens / air pollutant levels | Inform asthma action plans based on local conditions | Anticipate peak seasonal spikes in allergen and air pollutant levels | Inform transportation plans to shift road use away from "asthma corridors" |

### 2.0   Creating Intelligent Infrastructure: The Path Forward

We are far from having the agile infrastructure that exhibits the breadth of capabilities described above. Why? At a fundamental level, we are breaking into a new world where physical capabilities and services are intertwined with the growing power of computation, the same power that has transformed numerous business sectors from commerce to healthcare to financial services. Our current municipal infrastructure is reminiscent of stock brokers yelling and waving paper in the air and physicians pouring over paper, wielding manual tools.

The foundation for Intelligent Infrastructure draws from and requires basic research and continued advancements in:

**Cyber Physical Systems (CPS):** Cyber Physical Systems also known as the Internet-of-things (IoT), will provide the computational substrate that will connect our future infrastructure to intelligent systems and software. Future visions of IoT-enabled infrastructure predict seven trillion intelligent sensors for over seven billion people. Most of the seven trillion sensors, actuators, and embedded computing devices will give rise to a new form of computing that is attached to the physical world. This new era of computing is predicted to have tremendous economic impact; estimates are north of $14 trillion over the next decade. Therefore, investing in IoT-enabled intelligent infrastructure could not only result in a new form of computing, it could also be the most significant economic impact since the Internet revolution. Bridging the gap between existing infrastructure and future computing requires linking the physical world with the digital, merging computation with sensors and actuators, to bridge this fundamental gap. Intelligent infrastructure is like a robot turned inside out; each sensor, actuator and communication line coalescing into an intelligent system. Bridging this gap requires new capabilities to make sense of, and manipulate, the physical world that is the fabric of daily life. Because CPS connects software to the physical world, software errors can result in physical harm to people or property. Investments in intelligent infrastructure will cause greater integration between the cyber and physical worlds and, thus, greater collaboration is needed between both software developers and engineers to understand and design these systems to be safe and secure.

**Artificial Intelligence (AI) / Machine Learning / Data Analytics:** The "intelligence" of intelligent infrastructure stems from the ability to integrate information from a variety of distributed and disparate sources into discernable patterns, to relate those patterns to explanatory models, and to use those models to inform decisions. However, just capturing data is not enough—inferences must also be informed by physical, legal, and societal constraints. For example, traffic rules must be obeyed, critical services such as hospitals must be prioritized, and schools need to start on time. The models themselves must be transparent and human-interpretable so that the reason for actions or decisions can be understood and explained. Systems must also adapt their behavior to different user populations (e.g. citizens vs. government officials) and to circumstances (e.g. operating during a weather emergency vs. a normal workday vs. a holiday). Providing these capabilities while ensuring robustness, reliability, fairness, and consistency in the face of immense scale and diversity is both an opportunity and a continuing challenge for AI systems.

**Security, Safety and Privacy:** The pervasive collection, use, and dissemination of sensor information requires new advances in security and privacy, while the deployment and physical operation of systems that manipulate the physical world also require new advances in system safety protection. Each node or piece of an intelligent infrastructure must be secured, both in how it is physically constructed and in how it communicates to a network or other devices. These nodes must be physically safe as well, potentially withstanding environmental conditions and interacting with the general public. There is also a significant security risk of interconnecting all these previously disconnected infrastructures as it becomes a central point of attack. Finally, research in privacy-preserving approaches to data collection and use is needed given the likelihood of pervasive data collection across many fundamental aspects of daily life.

**Networking:** Intelligent infrastructure relies on an underlying communication network to collect and analyze data from distributed sensors, and adapt how the infrastructure operates in response. The network must offer low latency (to enable fast decisions), high reliability (to ensure the infrastructure operates seamlessly), and good security (to defend against cyberattacks). These networks also need to connect the smart infrastructure to cloud-computing resources that offer the necessary computational and storage resources to analyze and archive data, and combine multiple kinds of information from different sources (including other kinds of smart infrastructure). An open challenge is understanding when and what computation can occur at the edge of the network and when data should be transmitted to cloud computing capabilities.

**Systems Programming:** The sheer scale of infrastructure systems raises new software engineering challenges and demands new specification and programming models. When deploying a new application, we cannot hope to re-program individual sensors, actuators, vehicles, network switches, and databases. Rather, what is needed are new abstractions and new programming models that support programming at the level of the whole, instead of the parts, and software tools that transform these high-level requirements into programs suitable for the individual nodes. In turn, this scale will require new understanding of emerging security threats, new methodologies for debugging (both for correctness and performance), and new analytic tools for modeling whole system behaviors. Finally, to be cost effective, the underlying resources, such as the network or compute nodes, will need to be shared across many applications, the same way a laptop can simultaneously run many programs. This capability raises additional systems issues around resource coordination and security.

**Decision Support:** Intelligence does not reside solely in the technical infrastructure; it is jointly produced through amplifying and augmenting human abilities to discern, manage, and anticipate challenges in modern municipal systems and services. Future systems must help municipal administration and staff to accurately diagnose current and pending problems in infrastructure, to effectively compare and contrast possible plans, and to efficiently allocate resources across a constrained system.

**Citizen Engagement:** Productive interactions with the general public is also fundamental to the success of intelligent infrastructure and a key research gap is developing tools and techniques to support those interactions. Challenges include reliably collecting information from the public (e.g. person as sensor), facilitating trust in increased data collection through pervasive sensors and automated and semi-automated systems, and creating applications for dialogue and civic engagement for shared governance and democratic participation.

## 3.0    Challenges and Opportunities

A concerted national agenda for Intelligent Infrastructure offers many opportunities including improving global competitiveness, catalyzing workforce development and investing in rural communities while also posing commensurate challenges in ensuring the usability and sustainability of this 21st century infrastructure.

**Global Competitiveness**: Intelligent infrastructure is quickly becoming central to providing services ranging from water, to energy, to multi-modal transportation, to health, to communications. And, economic competitiveness is increasingly tied to the reliability and resilience of these critical systems[2]. Simply put, communities without robust intelligent infrastructure systems will be left behind in the global economy because their critical infrastructure systems — utilities, energy, transportation, health, and emergency services — will not be competitive compared to places who made the investments in cyber-physical systems to support operations. Failing to invest in intelligent infrastructure misses the opportunity to provide local communities with globally competitive roads, bridges, and transit but also abdicates the opportunity to build a new industry around the products, services, and systems developed on the platform of intelligent infrastructure

**Needs of Rural Communities:** The needs of metropolitan areas compared to rural communities vary tremendously and we recommend focused attention on the needs of small and rural communities in addition to addressing the needs of dense urban settings. Basic access to Internet-based capabilities is critical to enabling basic citizen engagement. A recent PCAST report[3] points to the pervasive access needs of aging adults, especially in small and rural communities. Hence, more research will need to be done on mobile platforms, mobile integrated end-to-end systems with easy setup, portable, low-cost, data cyber-infrastructures, edge computing and tele-services that allow for different economic contexts.

**Education and workforce development:** We wish to amplify the importance of educational programs and approaches that integrate key training in data analytics, sensing, communication, security, and privacy as deploying intelligent infrastructure will require greater ability and awareness of communities to manage, maintain, and protect the infrastructure[4]. We also want to call attention to the need for basic and applied research in workforce tools that will enable people to access and harness these capabilities. For example, research in visual analytics addresses challenges of working with complex data sets, understanding probabilistic and predictive information and supporting collaborative decision making. Likewise, wearable and augmented reality systems offer the ability to "see" and interact with layers of information connected to physical objects.

**Interoperability:** Intelligent infrastructure presents particular challenges because it is integrated both into and across different critical, existing infrastructures. From water and electricity systems and across built, natural, and socio-economic environments, robust intelligent infrastructure is increasingly required for the secure and resilient operations of government services and systems. Consequently, this infrastructure-of-infrastructures presents a unique problem for critical infrastructure: how to integrate the capabilities and capacities of intelligent infrastructure into incumbent systems while mitigating interruptions, reducing exposure to threats, and ensuring continuity of service? In short, intelligent infrastructure requires attention in its own right as a new critical public infrastructure.

---

[2] [1] See Clark, Jennifer (2013) <u>Working Regions: Reconnecting Innovation and Production in the Knowledge Economy</u>, London: Routledge. See also Clark, Jennifer, Hsin I. Huang, and John P. Walsh (2010) A *Typology of Innovation Districts: What it Means for Regional Resilience*. Cambridge Journal of Regions, Economy and Society. 3 (1): 121-137.
[3] President's Committee of Advisors on Science and Technology, Report to the President on the Independence, Technology, and Connection in Older Age, March 2016.
[4] *A 21st Century Cyber-Physical Systems Education*. National Academies Press. 2016. https://www.nae.edu/164520.aspx

**Sustainability:** Sustainability is a formidable barrier for the long-term success of intelligent infrastructure investments. These barriers may be especially high in the case of small towns and rural areas where resources are severely limited. First, local governments need concrete, actionable plans from vendors or groups of researchers who propose deploying intelligent infrastructure. Another major challenge of sustainability is lack of innovative economic models to deploy and upgrade intelligent infrastructures. Some gains (e.g., decreasing crime) may not have direct revenue implications while others (e.g., decreasing parking) may reduce city revenue. A final concern is that while infrastructure is upgraded on the time scale of decades, computing technology changes every year. Building "smart technology" into infrastructure requires designing the technology to be upgradable over time without replacement.

**Infrastructure for research and authentic evaluation:** We also wish to emphasize the importance of research infrastructure and "authentic evaluation," i.e., evaluating systems in the context of real use. Developing a comprehensive plan for investing in research infrastructure remains an ongoing challenge. Additionally, many evaluation metrics are non-traditional. Success may not be measured as to whether a technology is robust, secure, or real-time (traditional computing metrics), but rather whether its deployment increases the number of visitors, new residents, and business activity, or decreases crime, traffic, and waste. Hence, evaluation of intelligent infrastructure technologies must bring together teams of engineers, computer scientists, social scientists, urban planners, economists and local leaders.

## 4.0    Recommendations

There is a clear need to invest in our nation's infrastructure, the roadways, water mains, energy distribution systems and more that support the fabric of everyday life. At this critical juncture, we must envision the future of these systems, anticipating the future demand on these systems and the role they play in the near and long term economic vitality of our communities. These capabilities are transformational and their impact, from the merging of the cyber and physical world to the scale of data collection made possible, requires considerable leadership and innovation. The path forward calls for sustained investment and collaboration between local communities, research universities, and industry with a critical role for leadership and investment by the federal government. Only together, can we reinvigorate the infrastructure that fuels our nation's activities at a significant scale to propel us into the next century of innovation and economic growth.

**Importance of Federal Leadership and Investment**: The Federal Government has an important role to play in shaping the scope and scale of intelligent infrastructure investments going forward. Simply put, the Federal Government will decide the platform on which the national economy is built going forward and whether it meets 20th century standards or sets the standard for the 21st century. There is a significant amount of basic research required to achieve the promise of intelligent infrastructure. Some of that research can be resourced through programs like the Smart and Connected Communities program or the Critical Resilient Interdependent Infrastructure Systems and Processes (CRISP) program of the National Science Foundation. However, the current resources are modest investments in basic research and not of a sufficient scale to support the broad, national technology deployments necessary.

Because intelligent infrastructure technologies cut across domains they also do not fit neatly under a specific federal agency. The MetroLab Network is a network of 38 cities, 4 counties, and 51 universities, organized into "city (or county) – university partnerships" focused on "research, development, and deployment" (RD&D) projects that offer technologically- and analytically-based solutions for challenges facing communities: mobility, security and opportunity, aging infrastructure, and economic development.  One role for the Federal Government is in resourcing and institutionalizing these networked partnerships to support policy diffusion across communities and information exchange about how smart communities' investments (programs, projects, and objects) perform as implemented. These networks allow local governments to achieve some economies of scale, build capacity, and avoid replicating mistakes or reinventing the wheel.

Although the Federal Government should not set a standardized approach, the Federal Government should consider developing technical standards and platforms for data, connectivity, and integration of hard infrastructure and information and communication technologies to protect citizens and consumers from excessive experimentation. The National Transportation Safety Board's approach to guidance on autonomous vehicles is a good example of signaling to industry, local governments, and researchers about how to shape strategic planning and private investment while protecting consumers and citizens. The National Institute of Standards and Technology's efforts to develop the global cities team challenge and convene industry, local governments, and universities to discuss and develop standards is an important start as well.

**Importance of Community and Industry Partnerships:** While there is great commercial interest in "smart city" and "IoT" technologies, it is unreasonable to expect that private industry can shoulder the sole responsibility for developing these national capabilities. In fact, uncertainties in this market, the substantial need for pre-competitive research, and the lack of standards and interoperable platforms create significant barriers to commercial progress and global leadership. Similarly, cities, communities and municipalities lack the expertise and financial resources to lead the nation through a period of risky experimentation. However, what is increasingly important is charting the path for sustained public-private partnerships that drive problems and approaches from the priorities of US cities and communities and develop the foundation for sustainable business models.

**Importance of Research Universities in Charting the Path for Intelligent Infrastructure:** In the U.S., the national innovation system largely relies on publicly-funded basic research and development conducted within the network of world class research universities throughout the country. For decades, these universities have served as the research and development backbone of U.S. industry and of national defense[5]. Research indicates that this national innovation has been effective in bringing forward new technologies and in facilitating the commercialization of new products, processes, and materials.

Research universities have extensive experience partnering with industry and government on technology diffusion projects like intelligent infrastructure. Research universities are built to test new technologies, evaluate alternatives, assess investments, evaluate economic impacts, measure distributional consequences, and certify processes, materials, products, and standards. As with any new enabling technology, research universities can play a role as a neutral third party with specialized technical expertise. Further, universities are embedded in local communities and have long-term working relationships with local and state governments and a vested interest in the presence of world class infrastructure in their own communities.

Finally, as research universities train the next generation of workers, citizens, and entrepreneurs, it is important to recognize that living and working in communities infused with intelligent infrastructure will be distinct from the built environment in which we live now. Whether the changes are immediately disruptive like autonomous vehicles or incremental adjustments to the skills required for living in and navigating the built environment (think automated grocery store check-outs, smartphone based parking systems), investments in technical training for new and incumbent workers will be required to take advantage of the value-added these technologies bring to the labor market. Universities again will be critical partners in developing both these technologies and the skilled workforce required to capitalize on their contributions to national and regional growth.

*This material is based upon work supported by the National Science Foundation under Grant No. 1136993. Any opinions, findings, and conclusions or recommendations expressed in this material are those of the authors and do not necessarily reflect the views of the National Science Foundation.*

---

[5] Clark, Jennifer (2014) Citing "Scientific Spaces" in the US: The Push and Pull of Regional Development Strategies and National Innovation Policies. Environment and Planning C: Government and Policy. Pp.1-16